\documentclass{IEEEtran}
\usepackage[utf8]{inputenc}
\usepackage{graphicx,amsmath,amsfonts,enumitem,pbox,url}
\usepackage[linesnumbered,ruled]{algorithm2e}

\def\BibTeX{{\rm B\kern-.05em{\sc i\kern-.025em b}\kern-.08em
    T\kern-.1667em\lower.7ex\hbox{E}\kern-.125emX}}

\title{Adaptive Modular Exponentiation Methods v.s. Python's Power Function}
\author{Shiyu Ji, Kun Wan\\
\url{{shiyu, kun}@cs.ucsb.edu} \\
Department of Computer Science \\
University of California Santa Barbara}
\date{}

\begin{document}

\maketitle

\begin{abstract}
    In this paper we use Python to implement two efficient modular exponentiation methods: the adaptive $m$-ary method and the adaptive sliding-window method of window size $k$, where both $m$'s are adaptively chosen based on the length of exponent. We also conduct the benchmark for both methods. 
    Evaluation results show that compared to the industry-standard efficient implementations of modular power function in CPython and Pypy, our algorithms can reduce 1-5\% computing time for exponents with more than 3072 bits.
\end{abstract}

\newcommand{\pow}{\textsf{pow}}

\section{Introduction}
In cryptography, it is particularly important to compute exponentiation of large base and exponent efficiently. The standard of RSA recommends the primes should have bit-length of at least 2048, and modular exponentiation is intensively used in RSA encryption and decryption. In general large integers can give better security and it is worth researching on highly efficient modular exponentiation algorithms for cryptographic large integers.

In this paper we focus on the use of script languages like Python not only because more and more cryptographic libraries are implemented in script languages, but also since Python has one of the most efficient libraries of built-in scientific computing functions. Any improvement on the performance of the built-in functions in Python would be beneficial to numerous applications, especially in the fields such as cryptography, natural language processing and machine learning. Thus it is worth checking if there is any space to improve.

In Python 2, we implement two adaptive modular exponential methods: $m$-ary \cite{MO96,Koc95} and sliding window of size $m$ \cite{MO96,Koc92}. Both the algorithms can adaptively choose the parameter $m$ with the goal to minimize the number of multiplications between large integers. To argue that both the methods can be efficiently deployed up to industry standard, we choose CPython and Pypy's built-in power functions as the baselines for comparison. Our experiment results show that for large exponents (e.g., 4096-bit), our methods can reduce the baseline running time by about 5\%.

\section{Adaptive $m$-ary Method}
This section will present the motivation and details of our adaptive $m$-ary method.

\begin{algorithm}[!th]
\SetArgSty{textnormal}
\DontPrintSemicolon
Take as input the modulus $N$, the base $g$ and the exponent $e$, and we are supposed to compute $g^e \mod N$\;
Parse $e$ as $(e_{t} e_{t-1} \cdots e_1 e_0)$ where each $e_i$ takes $m$ bits\;
$g_0 \gets 1$\;
\For{$i$ from 2 to $2^m-1$}{
    $g_i \gets g_{i-1}\cdot g \mod N$\;
}
$A \gets g_{e_{t}}$\;
\For{$i$ from $t-1$ down to 0}{
    \For{$j$ from 1 to $m$}{
        $A \gets A\cdot A \mod N$\;
    }
    $A \gets A\cdot g_{e_i} \mod N$\;
}
Return $A$.
\caption{Left-to-right $m$-ary exponentiation}
\label{alg:lrmary}
\end{algorithm}

Recall the left-to-right $m$-ary exponentiation (14.82 in \cite{MO96}) as Algorithm \ref{alg:lrmary}. 
In the precomputation (lines 3 to 6), we need $2^m-2$ multiplications to compute $g^2$, $\cdots$, $g^{2^m-1}$.
In the iteration from line 7 to 11, we have $tm = k-m$ times of squaring (line 10), where $k = (t+1)m$ is the bit-length of exponent $e$, and in average $(1-2^{-m})t$ times of multiplication on line 12. Hence in average the number of multiplications is given as follows.
$$T(k, m) = 2^m - 2 + k - m + (1-2^{-m})\frac{k-m}{m}.$$
The reasoning above follows the similar lines given by \cite{Koc92,Koc95}, which also presented a similar equation.

An interesting fact of $T(k, m)$ is that there always exists an positive integer $m^*$ that minimizes $T(k, m)$ for any positive integer $k$, since $T(k, m)$ is convex on $m$ for any $k$ (see Appendix for the proof). Moreover, for any $k$ the minimizer $m^*$ satisfies $T(k, m^*) \leq T(k, m^* + 1)$ by convexity of $T(k, m)$.
It turns out the minimizers $m^*$ are usually small, i.e., less than 10. Hence it is easy to find $m^*$ given any $k$. Table \ref{tab:minis} gives the minimizers for different $k$'s.

\begin{table}[!th]
    \centering
    \begin{tabular}{|c||c|c|c|c|}
    \hline
        $k$ & 1-5 & 6-34 & 35-121 & 122-368 \\
        \hline
        $m^*$ & 1 & 2 & 3 & 4 \\
        \hline
    \hline
        $k$ & 369-1043 & 1044-2822 & 2823-7370 & 7371- \\
        \hline
        $m^*$ & 5 & 6 & 7 & 8 \\
        \hline
    \end{tabular}
    \vspace{1em}
    \caption{The minimizers $m^*$ given bit-length $k$ of the exponent.}
    \label{tab:minis}
\end{table}

\newcommand{\ary}{\textsf{aryExp}}
Since multiplication between large integers with thousands of bits is expensive, we leverage the minimizers to reduce the multiplication times as many as possible. This gives rise to our adaptive $m$-ary method. Algorithm \ref{alg:adapt} gives the detailed procedure, where $m$-\ary($g$, $e$, $N$) denotes the $m$-ary method to compute $g^e \mod N$.

\begin{algorithm}[!th]
\SetArgSty{textnormal}
\DontPrintSemicolon
Take as input the modulus $N$, the base $g$ and the exponent $e$, and we are supposed to compute $g^e$\;
Let $k$ be the bit-length of $e$\;
If $k < 6$, return 1-\ary($g$, $e$, $N$)\; 
If $k < 35$, return 2-\ary($g$, $e$, $N$)\; 
If $k < 122$, return 3-\ary($g$, $e$, $N$)\; 
If $k < 369$, return 4-\ary($g$, $e$, $N$)\; 
If $k < 1044$, return 5-\ary($g$, $e$, $N$)\; 
If $k < 2823$, return 6-\ary($g$, $e$, $N$)\; 
If $k < 7371$, return 7-\ary($g$, $e$, $N$)\; 
return 8-\ary($g$, $e$, $N$).
\caption{Adaptive $m$-ary exponentiation}
\label{alg:adapt}
\end{algorithm}

\section{Adaptive Sliding-Window Method of Window Size $m$}
This section will present the motivation and details of our adaptive sliding-window method of window size $m$.

\begin{algorithm}[!th]
\SetArgSty{textnormal}
\DontPrintSemicolon
Take as input the modulus $N$, the base $g$ and the exponent $e$, and we are supposed to compute $g^e \mod N$\;
Parse $e$ as $(e_{t} e_{t-1} \cdots e_1 e_0)$ where each $e_i$ is one bit\;
$g_1 \gets g \mod N$, $g_2 \gets g^2 \mod N$\;
\For{$i$ from 1 to $2^{m-1}-1$}{
    $g_{2i+1} \gets g_{2i-1}\cdot g_2 \mod N$\;
}
$A \gets 1$, $i \gets t$\;
\While{$i \geq 0$}{
    \If {$e_i= 0 $}{$A\gets A\cdot A$, $i \gets i-1$\;}
    {
        Find the longest bitstring $e_i e_{i-1} \cdots e_\ell$ s.t. $i-\ell+1\leq m$ and $e_\ell = 1$\;
        \For{$j$ from 1 to $i-\ell+1$}{$A \gets A\cdot A$\;}
        $A \gets A \cdot g_{e_i e_{i-1}\cdots e_\ell}$\;
        $i \gets \ell-1$\;
    }
}
Return $A$.
\caption{Left-to-right sliding-window exponentiation with window size $m$}
\label{alg:lrwin}
\end{algorithm}

Recall the left-to-right sliding-window exponentiation of window size $m$ (14.85 in \cite{MO96}) as Algorithm \ref{alg:lrwin}. 
In the precomputation (lines 3 to 6), we need $2^{m-1}$ multiplications to compute all the necessary powers of $g$.
In the iteration from line 9 to 14, we have $k$ times of squaring (lines 10 and 15), where $k = t+1$ is the bit-length of exponent $e$, and in average $k/(m+1)$ times of multiplication on line 16 (the mean length of one window with its forthcoming zeros is $m+1$). Hence in average the number of multiplications is given as follows.
$$T'(k, m) = 2^{m-1} + k + \frac{k}{m+1}.$$

One can similarly verify that there always exists an positive integer $m^*$ that minimizes $T'(k, m)$ for any positive integer $k$, since $T'(k, m)$ is also convex on $m$ for any $k$ (see Appendix). Table \ref{tab:minis_win} gives the minimizers for different $k$'s.

\begin{table}[!th]
    \centering
    \begin{tabular}{|c||c|c|c|c|}
    \hline
        $k$ & 1-20 & 21-23 & 24-79 & 80-239 \\
        \hline
        $m^*$ & 2 & 3 & 4 & 5 \\
        \hline
    \hline
        $k$ & 240-671 & 672-1791 & 1792-4607 & 4608-11519 \\
        \hline
        $m^*$ & 6 & 7 & 8 & 9 \\
        \hline
    \end{tabular}
    \vspace{1em}
    \caption{The minimizers $m^*$ given bit-length $k$ of the exponent.}
    \label{tab:minis_win}
\end{table}

\newcommand{\win}{\textsf{winExp}}

Similarly as Algorithm \ref{alg:adapt}, Algorithm \ref{alg:adapt_win} gives the detailed procedure of adaptive sliding-window exponentiation method.
In Algorithm \ref{alg:adapt_win} $i$-\win($g$, $e$, $N$) denotes the sliding window method with window size $i$ to compute $g^e\mod N$.

\begin{algorithm}[!th]
\SetArgSty{textnormal}
\DontPrintSemicolon
Take as input the modulus $N$, the base $g$ and the exponent $e$, and we are supposed to compute $g^e$\;
Let $k$ be the bit-length of $e$\;
If $k < 21$, return 2-\win($g$, $e$, $N$)\; 
If $k < 24$, return 3-\win($g$, $e$, $N$)\; 
If $k < 80$, return 4-\win($g$, $e$, $N$)\; 
If $k < 240$, return 5-\win($g$, $e$, $N$)\; 
If $k < 672$, return 6-\win($g$, $e$, $N$)\; 
If $k < 1792$, return 7-\win($g$, $e$, $N$)\; 
If $k < 4608$, return 8-\win($g$, $e$, $N$)\; 
return 9-\win($g$, $e$, $N$).
\caption{Adaptive sliding-window exponentiation with window size $m$}
\label{alg:adapt_win}
\end{algorithm}

\section{Evaluation}

\begin{table*}[!th]
    \begin{tabular}{|c||c|c|c|c|c|}
    \hline
        \#bits $k$ & CPython $\pow$ (ms) & Adaptive $m$-ary (ms) & Change ratio of $m$-ary & Adaptive sliding window (ms) & Change ratio of sliding window \\
        \hline
        1024 & 5.050 $\pm$ 0.84 & 5.271 $\pm$ 0.90 & +4.39\% & 5.179 $\pm$ 0.83 & +2.56\% \\
        2048 & 32.942 $\pm$ 2.81 & 32.88 $\pm$ 2.04 & -0.17\% & 32.41 $\pm$ 2.69 & -1.63\% \\
        3072 & 101.136 $\pm$ 6.35 & 99.77 $\pm$ 5.82 & -1.35\% & 97.273 $\pm$ 6.33 & -3.82\% \\
        4096 & 229.435 $\pm$ 11.48 & 224.432 $\pm$ 10.27 & -2.31\% & 218.00 $\pm$ 10.67 & -4.98\% \\
        \hline
    \end{tabular}
    \vspace{1em}
    \caption{Time cost of adaptive $m$-ary and sliding window v.s. CPython $\pow$.}
    \label{tab:res}
\end{table*}

\begin{table*}[!th]
    \begin{tabular}{|c||c|c|c|c|c|}
    \hline
        \#bits $k$ & Pypy $\pow$ (ms) & Adaptive $m$-ary (ms) & Change ratio of $m$-ary & Adaptive sliding window (ms) & Change ratio of sliding window \\
        \hline
        1024 & 3.450 $\pm$ 0.19 & 3.586 $\pm$ 0.73 & +3.94\% & 3.543 $\pm$ 0.95 & +2.69\% \\
        2048 & 20.035 $\pm$ 2.08 & 20.011 $\pm$ 2.28 & -0.12\% & 19.731 $\pm$ 2.18 & -1.52\% \\
        3072 & 58.383 $\pm$ 5.61 & 57.782 $\pm$ 6.99 & -1.03\% & 56.118 $\pm$ 5.32 & -3.88\% \\
        4096 & 129.20 $\pm$ 10.94 & 126.556 $\pm$ 11.82 & -2.04\% & 123.14 $\pm$ 10.07 & -4.69\% \\
        \hline
    \end{tabular}
    \vspace{1em}
    \caption{Time cost of adaptive $m$-ary and sliding window v.s. Pypy $\pow$.}
    \label{tab:res_pypy}
\end{table*}

This section presents the evaluation results of our adaptive $m$-ary and $m$-sized sliding window method, with the baseline power functions from the popular implementations, CPython and Pypy, which are considered to be very efficient in practice \cite{RO15, Py06}.

\subsection{Experiment Setup}
To test the performance of our algorithm and CPython/Pypy's power functions, we choose at uniformly random the base $g$, exponent $e$ and modulus $N$ to be large integers that are sufficient for cryptographic use, i.e., with bit-length of 1024, 2048, 3072 and 4096. We guarantee that $N$ is larger than $g$ and $e$.
For exponentiation of each bit-length, we take 1000 samples and compute the average computing time. 
We use Python 2 with CPython/Pypy as the implementation.
We run our Python code on a Linux Ubuntu 16.04 server with 8 cores of 2.4 GHz AMD FX8320, 16GB memory.

\subsection{The Baseline}
The power function $\pow(g, e, N)$ implemented in CPython \cite{longpow17} works as follows. For exponents with more than 8 digits, $\pow$ uses 5-ary method. For exponents of no more than 8 digits, $\pow$ uses LR binary method as Algorithm \ref{alg:lrmary}.
Pypy's $\pow$ always uses LR binary method \cite{powPypy17}.
To maximize the efficiency, $\pow$ as well as most code in CPython is written in C language, and Pypy has more efficient (on average 7.5 times faster than CPython) arithmetic operations including multiplication \cite{speedPy17}. Note that our experiment code is written in Python 2, implying there is still some improvement space if we write our algorithm in C language.

\subsection{The Performance Results}

Table \ref{tab:res} gives the testing results of the three methods implemented in CPython, and Table \ref{tab:res_pypy} gives the results of the methods implemented in Pypy. The times are measured in milliseconds, and the errors are the sampled deviations.
We first discuss the CPython results. Note that for exponents with more than 2048 bits, our adaptive $m$-ary and $m$-sized sliding window method overall outperform CPython and Pypy's $\pow$ implementation, since for such large exponents, the minimizer $m^*$ is often more than 5, and thus the strategy in CPython or Pypy's $\pow$ is sub-optimal, whereas our adaptive method still captures the minimizer. 
For small exponents CPython and Pypy's $\pow$ outperforms ours, since CPython $\pow$'s C implementation is more efficient than our Python 2 code, and Pypy $\pow$'s 1-ary LR method does not need to precompute or memorize any powers. In particular, for $1024$-bit exponent, both $\pow$ and our $m$-ary method choose $m = 5$, and thus run the same algorithm. Hence the $+4.39\%$ change is mainly contributed by the difference between C and Python 2. In general, adaptive sliding window method achieves the best performance among the three methods. In particular, for 4096-bit modular exponentiation, adaptive sliding window method can reduce $\pow$'s time by nearly 5\%.

For Pypy's results, our algorithms achieve similar performance gains, e.g, 4.6\% for 4096-bit exponent.

\subsection{Discussion on the Exponentiation with Short Exponents}

Our experiments above indicate that for relatively short exponents (e.g., less than 1024 bits), the python built-in power function outperforms ours, due to the aforementioned gap on efficiency between Python and the low-level language for implementation like C. To mitigate this problem we may treat the short exponents differently, e.g., if the length of the exponent is no more than 1024, we just use Python's built-in function to calculate the power. In this way our performance should be competitive for short exponents, while achieves better for long exponents. 

\section{Conclusion}
We have presented two improved modular exponential algorithms based on $m$-ary and sliding window respectively. 
Our methods can adaptively choose $m$ based on the length of exponent.
To verify the improvement on performance, we have done the benchmark by comparing their time cost with the power function implemented in CPython/Pypy as a baseline.
The comparison results have verified that our methods outperform for very large exponent, e.g., 4.6-5\% reduction on time for exponents with 4096 bits.

\bibliographystyle{plain}
\bibliography{references}

\appendix
We will prove the claim that $T(k, m)$ is convex on $m\in\mathbb{R}^+$ for any $k > m$. It suffices to show $\frac{\partial T(k, m)}{\partial m}$ monotonically increases on $m\in\mathbb{R}^+$ for any $k > m$.
Compute the derivative as
$$\begin{aligned}
\frac{\partial T(k, m)}{\partial m} = &(k-m)\left(\frac{2^{-m}\log 2}{m} - \frac{1-2^{-m}}{m^2}\right) \\
&- \frac{1-2^{-m}}{m} + 2^m \log 2 - 1.
\end{aligned}$$
It is routine to verify the monotonic increase of $- \frac{1-2^{-m}}{m}$ and $2^m \log 2$, by verifying their derivatives are always positive over $\mathbb{R}^+$.

It remains to show that the term
$$\begin{aligned}
R(k, m) = &\frac{2^{-m}\log 2}{m}- \frac{1-2^{-m}}{m^2} \\
=& m^{-2}((m\log 2+1)\cdot 2^{-m}-1)
\end{aligned}$$
is always negative over $\mathbb{R}^+$ and monotonically increases on $m$ for any $k>m$.

Since $1+m\log 2 < 2^m$ (by Taylor's theorem) for any positive $m$, $R(k, m)$ is always negative over $\mathbb{R}^+$.
To verify the monotonic increase, compute the derivative
$$\begin{aligned}
\frac{\partial R(k, m)}{\partial m} = &2^{-m}m^{-3}(-m^2\log^2 2 \\&- 2m\log 2 - 2 +2^{m+1}).
\end{aligned}$$
Since $2^{m+1} > 2 + 2m\log 2 + m^2 \log^2 2$ (again, by Taylor's theorem), 
$$\begin{aligned}
\frac{\partial R(k, m)}{\partial m} > 0.
\end{aligned}$$
Hence for any $k>m$, $R(k, m)$ monotonically increases over $m\in\mathbb{R}^+$.

The function 
$$T'(k, m) = 2^{m-1} + k + \frac{k}{m+1}$$
is also convex over $m \in \mathbb{R}^+$ for any positive $k$, since
$$\frac{\partial^2 T'(k, m)}{\partial m^2} = 2^{m-1}\cdot \log^2 2 + \frac{2k}{(m+1)^3}$$
is always positive over $m \in \mathbb{R}^+$.

\end{document}